# On the effect of Ti on Oxidation Behaviour of a Polycrystalline Nickel-based Superalloy

S. Pedrazzini[1,2*], B. S. Rowlands[2], A. Turk[2], I. M.D. Parr[3], M. C. Hardy[3], P. A. J. Bagot[4], M. P Moody[4], E. Galindo-Nava[2], H. J. Stone[2]

[1]Department of Materials, Faculty of Engineering, Imperial College London, South Kensington Campus, Exhibition Road, London, SW7 0FS, UK
[2]Department of Materials Science and Metallurgy, University of Cambridge, 27 Charles Babbage Road, Cambridge, CB3 0FS, UK
[3] Rolls-Royce plc, PO Box 31, Derby, DE24 8BJ, UK
[4] Department of Materials, University of Oxford, Parks Road, Oxford, OX1 3PH, UK

*Corresponding author: s.pedrazzini@imperial.ac.uk

**Abstract:**
Titanium is commonly added to nickel superalloys but has a well-documented detrimental effect on oxidation resistance. The present work constitutes the first atomistic-scale quantitative measurements of grain boundary and bulk compositions in the oxide scale of a current generation polycrystalline nickel superalloy performed through atom probe tomography. Titanium was found to be particularly detrimental to oxide scale growth through grain boundary diffusion.

Ni-based superalloys can be exposed to temperatures approaching 75% of the melting point in service, causing formation of oxide scales [1]. This oxidation causes both material loss (spallation) and the deterioration of the mechanical properties through the depletion of alloying elements in the oxide-affected zone [2,3]. Ni-based superalloys have outstanding oxidation resistance due to the formation of scales of $Cr_2O_3$ or $Al_2O_3$, which passivate the surface preventing further oxidation [4]. However, the oxidation of such complex, multi-component alloys can be strongly influenced by minor alloying addition, allowing considerable scope for optimisation. Ti is a common addition that strengthens the coherent intermetallic precipitates γ' [5] to which Ni-based superalloys owe their outstanding mechanical properties. However, increased Ti concentrations exacerbate the rate of oxidation [6].

Oxide characterisation studies are typically performed using SEM-EDX and XRD [6–12]. Studies of oxidised alloys using high-resolution characterisation techniques have, until recently, been limited. Techniques such as Atom Probe Tomography (APT) have the spatial and chemical resolution required to characterise the oxide scales and grain boundaries within them. Recent studies performed using high-resolution characterisation techniques have shown that thermally grown oxides are nano-crystalline and contain nano-scale phases which cannot be detected using low resolution characterisation techniques such as SEM-EDX and XRD, or bulk techniques like EXAFS [13–19]. Nano-sized grains have higher surface-to-volume ratios, therefore due to the extensive amount of grain boundary area, grain boundary diffusion is expected to be dominant. In the present study, the surface oxide scale on superalloy RR1000 was characterised using high-resolution techniques for the first time, giving new insights in the oxidation mechanisms operating in the alloy.

The oxidation behaviour of alloy RR1000 was studied previously [7,9–11] using Thermo-Gravimetric Analysis (TGA), Scanning Electron Microscopy coupled with Energy Dispersive X-ray spectroscopy (SEM-EDX) and X-ray diffraction (XRD), the oxide scale formed on RR1000 was assessed after air exposures at 800°C for up to 5000 hours [8]. The surface oxide scale was Ti-doped chromia with rutile forming on its outer surface. This study reported sub-parabolic thickening kinetics for the Ti-doped chromia scale, with initial rates ~2 orders of magnitude higher than that of undoped (Ti-free) chromia. The suggested interpretation was that the Ti in solid solution acted as a higher-valence dopant ($Ti^{4+}$, which substituted for $Cr^{3+}$ in the crystal lattice), causing minimal lattice distortion but increasing the amount of Cr vacancies required to maintain overall charge neutrality. This possibility was considered not just by Cruchley *et al.* [8], but was also extensively discussed by Blacklocks *et al.* [20] in the context of Ti-doped chromia used in sensors for trace gases. Blacklocks *et al.* demonstrated using synchrotron-based extended X-ray absorption fine structures (EXAFS) that Ti is present as $Ti^{4+}$ ions, and that these substitute predominantly on the Cr lattice site, creating an excess of Cr vacancies [20]. Cruchley *et al.* [8] also commented that the enhanced diffusion kinetics and scale thickening rate created through Ti doping must apply to the oxide grain boundaries, since these are likely to remain the main diffusion routes at 700-800 °C.

Critically, the relative concentrations of Ti in the bulk oxide scale compared to the grain boundaries of the oxide has yet to be analysed. Given the importance of the relative contributions of grain boundary and bulk diffusion across the oxide scale caused by Ti doping, high-resolution characterisation studies are required to understand the mechanism behind the deleterious effect of Ti on oxidation resistance of nickel-based superalloys.

A sample of the polycrystalline disc alloy RR1000, approximately 5 mm x 5 mm x 7 mm, was supplied by Rolls-Royce plc (produced through powder metallurgy, hot isostatic pressing and isothermal forging). The nominal composition of the alloy is shown in Table 1.

| RR1000 | Ni | Cr | Co | Mo | Al | Ti | Ta | Hf | Zr | C | B | Si |
|---|---|---|---|---|---|---|---|---|---|---|---|---|
| Wt.% | Bal. | 15 | 18.5 | 5 | 3 | 3.6 | 2 | 0.5 | 0.06 | 0.03 | 0.015 | 0.5 |

*Table 1: nominal composition of the RR1000 alloy used in the present study.*

The sample was homogenised and heat-treated below the γ' solvus temperature in order to achieve a fine-grained microstructure. Surfaces were manually ground with 1200 grit SiC paper, then oxidised in a furnace at 800°C for 500 hours in air. The specimen was air-cooled, cut and mounted in conductive epoxy resin and polished using a protocol culminating in a 0.04 µm oxide nanoparticle solution for SEM analysis.

The oxide scale was imaged using a Zeiss GeminiSEM300 equipped with an Oxford Instruments EDX detector. Images and EDX compositional maps were taken using a 10 kV accelerating voltage, 3 nA beam current and working distance of 10 mm. EDX data were processed using Aztec software. Peak overlaps, such as Cr-L and O-K, were deconvoluted either using the Aztec TruMap algorithm or by manually selecting higher order peaks for each element. A representative SEM micrograph is shown in Figure 1, along with EDX elemental composition maps (data obtained from K-lines are shown for every element except for Ta, where the L-line was used).

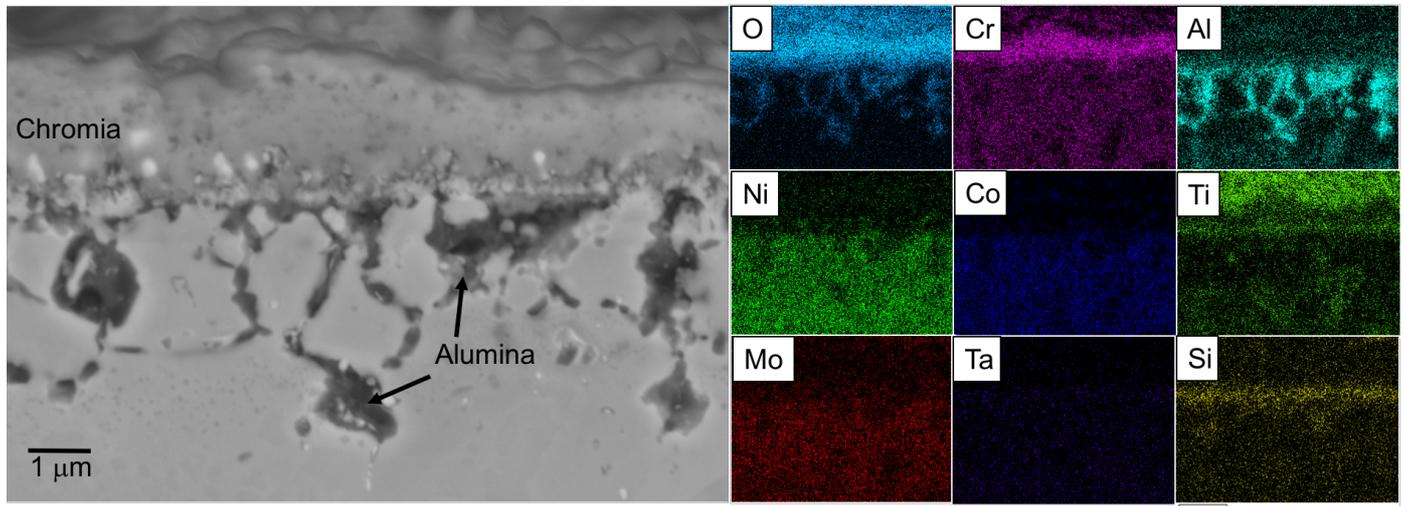

*Figure 1:* Backscattered electron micrograph of a RR1000 sample heat treated at 800 °C for 500 hours in air and EDX elemental composition maps showing a continuous layer of chromia, with a titanium rich layer on the outer surface (less visible in the backscattered image due to a lack of Z-contrast between Cr and Ti) and some discontinuous precipitation of sub-surface alumina. EDX data were obtained from the K-lines, except for Ta where the L line was used.

SEM micrographs provide an overview of the oxide scale and the phases present. Chromia and rutile are identified as part of the oxide scale, with a discontinuous layer containing elongated intergranular and intragranular alumina particles directly underneath. From SEM-EDX alone, the difference between the grain boundary and bulk Ti content within the chromia scale cannot be observed. The oxide scale thickness and alumina "finger" depth were measured from SEM micrographs. Five measurements per image were performed, using ImageJ software over 10 images. The mean thickness of the surface scale (chromia + rutile) was 2.0 ± 0.2 µm, while the alumina depth was 4.3 ± 1.4 µm adding up to a total oxidation damage depth of 6.3 µm.

Atom probe tomography (APT) samples were lifted out from the oxide scale using a FEI Helios dual beam SEM-FIB microscope and mounted onto Cameca silicon flat top coupons. Samples were protected from ion beam damage through the in-situ deposition of a Pt layer, before a wedge-shaped cantilever beam was extracted using an omniprobe micromanipulator. The cantilever was sectioned and mounted onto the silicon coupon using Pt welds. The samples were then sharpened until tips were <100 nm in diameter. Samples were inserted in a Cameca Local Electrode Atom Probe (LEAP) 5000 XR and analysed using a UV laser with 300 pJ of energy, with a pulse frequency of 200 kHz and a stage temperature of 55 K. Two samples were analysed: one from the surface rutile and one from the underlying chromia scale. Atomic scale reconstructions of both samples were performed using IVAS 3.6.12 software (Integrated Visualisation and Analysis Software) and are shown in Figure 2.

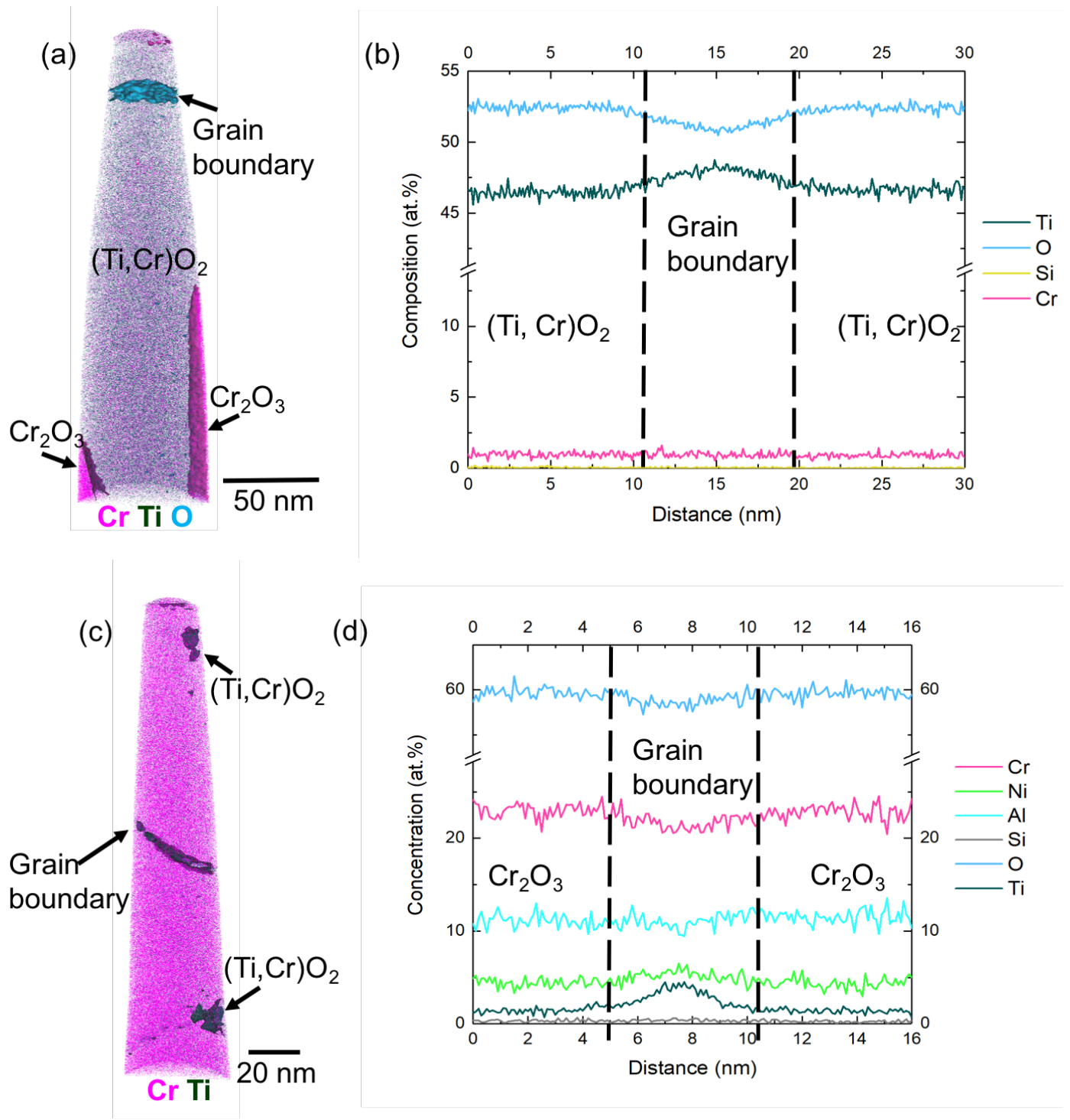

*Figure 2: (a) atom probe tomography reconstruction of the rutile growing on the outer surface of the oxide scale, showing some chromia inclusions and a grain boundary of a RR1000 sample heat treated at 800°C for 500 hours in air. Iso-concentration surfaces at 53 at% O and 10 at% Cr are shown. (b) 1D concentration profile across the grain boundary, highlighting the increase in Ti concentration and relative oxygen depletion present. (c) atom probe reconstruction of a chromia sample, containing nano-scale rutile inclusions and a grain boundary. Iso-Concentration surfaces at 3 at.% Ti are shown. (d) 1D concentration profile across the grain boundary, also showing a 4 at% Ti enrichment.*

APT is one of the few techniques that provide accurate, quantitative data on the atomistic scale including at grain boundaries. Figure 2 (a) shows an atom map of the rutile layer (from the outer surface of the oxide scale), containing chromia inclusions and a grain boundary. Figure 2 (b) shows the elemental composition across this grain boundary. A local enrichment of ~4 at.% Ti is observed at the grain boundary and a corresponding oxygen depletion, when compared to the value in the chromia. The presence of ~2 at.% Cr in solid solution within the rutile scale was also confirmed. Figure 2 (c) shows an atom map of the chromia (bulk of the oxide scale), again with a grain boundary and inclusions of rutile. A 1D concentration profile across the grain boundary is shown in Figure 2 (d). The Ti concentration in the bulk chromia is ~1 at.%. In addition, ~5 %at. Ni and 11 at.% Al were detected in solid solution in the chromia. These values match those measured by Kitaguchi *et al.* [13]. The oxygen concentration is stoichiometric in $Cr_2O_3$ (60%), though in the rutile part of the oxide scale (Figure 2 (b)) the content is sub-stoichiometric. This is likely due to the peak overlap between $O_2$ and TiO at 32 Da. A further study of running conditions and their effect on the stoichiometry, analysis conditions and peak overlap deconvolution is available in other publications [21]. The presence of rutile in the oxide scale was confirmed through XRD (see supplementary information) and for further analyses the oxide composition was accurately measured by using isotopic abundances to deconvolute the contribution of TiO and $O_2$ to the peak.

It is reported in the literature that at high oxygen partial pressure ($PO_2$), > $10^{-5}$ atm, $Cr_2O_3$ is a known p-type semiconductor with electron holes and chromium vacancies as the dominant defects [22]. This behaviour will be dominant close to the oxide surface, in direct contact with air. At intermediate $PO_2$, ~$10^{-5}$ – $10^{-14}$ atm, $Cr_2O_3$ behaves as an intrinsic semiconductor with electrons and electron holes being the dominant defects; and at low $PO_2$, <$10^{-14}$ atm, near the metal/$Cr_2O_3$ equilibrium oxygen pressure, it changes to an n-type semiconductor with electrons and chromium interstitials being dominant [22]. $Cr_2O_3$ can grow either through the diffusion of positive metallic ions outwards from the metal-oxide interface or through the inward diffusion of negatively charged oxygen ions and its predominant growth mechanism is dependent on the ionic conductivity, which, in turn, changes depending on the oxygen partial pressure [23,24]. Doping with ions of different valence states can change the ionic conductivity regime of the oxide scale, therefore affecting its growth mechanism [25]. This was shown previously by Holt and Kofstad [25] in the case of Mg (II) doped $Cr_2O_3$ at different $PO_2$. Similarly, α-alumina, which grows at very low partial pressure underneath the chromia scale, is an n-type ionic conductor and grows through the inward diffusion of negatively charged $O^{2-}$ ions towards the oxide/metal interface [26]. Table 2 summarises the preferred relative valence of each cation if they were found in solid solution in p-type $Cr_2O_3$ (towards the outer surface). All values of preferred cation valence are taken from the literature, taking into account the conditions of the present experimental setup. Samples were oxidised for 500 hours, which will have enabled the formation of stable oxides eliminating potential kinetic effects. The metastable initial oxides that formed are shown in the XRD experiment available as supplementary information. The predicted effect of each cation in solid solution on the relative oxidation rate is based on the assumption of electroneutrality. The relative increase of positive cations will have to be compensated by the increased uptake of negatively charged defects, leading to faster diffusion of ions, through the oxide scale.

| Cation valence | 2 | 3 | 4 | 5 |
|---|---|---|---|---|
| Stable oxides [1] | NiO [24] | $Al_2O_3$ [25] | $SiO_2$ [27] | $Ta_2O_5$ [26] |
|  |  | $Cr_2O_3$ [27] | $TiO_2$ [7] |  |
| Effective Valence of substitutional solutes in p-type $Cr_2O_3$ | -1 | 0 | 1 | 2 |
| Relative oxidation rate for p-type chromia | slower | neutral | fast | faster |

*Table 2: summary of cation valences and their relative effect in solid solution in p-type chromia. The effect would be reversed in the case of n-type chromia (close to the oxide-metal interface).*

The measured chemical composition of the chromia scale in this alloy (by APT) was 23 at.% Cr, 11 at.% Al, 5 at.%Ni, 1 at.%Ti. So, if the Cr and Al content are assumed to have no effect on the defect concentration, the remaining Ni and Ti will have opposite effects, though the higher Ni concentration will result in a net reduction in the concentration of negative defects, causing slower inwards diffusion of oxygen and suggesting that the oxide scale grows primarily through outwards diffusion of metallic cations. Atom probe tomography cannot be used to measure vacancy concentrations, the defect concentrations quoted in the present work are calculated from the concentrations of individual elements in the oxide scale.

Critically, as the atom probe showed, throughout the entire oxide scale Ti segregates primarily to grain boundaries within the $Cr_2O_3$ scale and to the rutile phase located on the outer surface. The deleterious effect of Ti on oxide scale thickness appears therefore to be directly correlated to outward diffusion of Ti from the bulk alloy, through the grain boundaries in the chromia scale, to the outer surface where it oxidises to form rutile.

A simple 2D diffusion calculation based on Fick's second law was used to estimate the outward flux of Ti through the grain boundaries. The aim was to demonstrate the effect of grain boundary diffusion on the magnitude of the flux available for further growth in a polycrystalline oxide scale compared to the single crystal scale, and not to compare directly the scale thickening rates in Ti-containing and Ti-free RR1000, as this would require producing a new alloy with identical conditions as RR1000 but without Ti. The scale growth itself was not included in the calculations nor was the effect of multicomponent diffusion – this is a challenging problem in itself and it is beyond the scope of this paper. Note however, that literature on oxidation modelling is largely focused on oxygen ingress [28] and subsurface elemental depletion [28,29], but not outward diffusion of alloying elements [28]. Additionally, existing models only consider the effective diffusion of a single element, usually oxygen [28,29]. Since the scale growth rate is related to the effective diffusivity of the elements contributing to its formation [28], the present calculations give a semi-quantitative estimate of the thickening rate as a function of grain boundary diffusion. The diffusivity and solubility used were taken from the NIMS diffusion coefficient database (Kakusan©). No data on Ti diffusivity in chromia were found, although the diffusivities of the other first-row transition metals are very similar. Mn was therefore chosen as the closest approximation. The diffusivities used were:

- Ti in Cr [30] = $2.32 \times 10^{-16}$ $m^2 s^{-1}$
- Mn in $Cr_2O_3$ [31] = $3.37 \times 10^{-21}$ $m^2 s^{-1}$
- Mn in $Cr_2O_3$ GB [31] = $9.14 \times 10^{-16}$ $m^2 s^{-1}$.

The relative solubilities of Ti were based on the APT measurements (Figure 2) assuming thermodynamic equilibrium. The Ti concentration in the bulk metal matrix directly underneath the chromia scale was obtained by SEM-EDX (1 at.%). The chromia scale was ~2 μm thick (measured by SEM) and its grain size was (mean planar diameter) ~300 nm, with 10 nm wide grain boundaries (estimated from APT data). The bulk metal is the supply of Ti, with the initial concentration corresponds to the total Ti concentration in the alloy (~20 mol $m^{-3}$). A zero-concentration boundary condition is present at the outer surface (because the sample ends), but periodic boundary conditions are assumed perpendicular to the oxide scale growth. The Ti flux peaked at the beginning (when the Ti concentration gradient was highest) then tapered off as the gradient decreased. At peak flux, the thickening rate in the polycrystalline scale was $3.77 \times 10^{-12}$ $m\ s^{-1}$, adding ~ 0.3 μm of rutile in 24 hours. In contrast, the single-crystal scale would be practically impermeable to Ti. The calculations results demonstrate that the outward Ti diffusion is almost entirely due to the presence of grain boundaries.

Our work demonstrates that atomistic-resolution characterisation of oxide scales can be coupled with diffusion modelling methods, to predict oxidation rates in complex multi-component alloys, but more work is required to obtain relevant thermo-kinetic growth parameters. Nonetheless, this opens new avenues for improved oxidation lifing models of engineering components.

**Acknowledgements:**

The authors would like to thank Rolls-Royce plc for the materials provision. The Oxford Atom Probe facility is funded by EPSRC (EP/M022803/1). This work was sponsored by the Rolls-Royce – EPSRC strategic partnership grant EP/M005607/1. Dr Galindo-Nava thanks the RAEng for their support by means of a research fellowship.